\documentstyle[preprint,aps,epsf]{revtex}

\begin{document}

\draft

\title{Chirality effects in carbon nanotubes.}

\author{E.L.~Ivchenko}

\address{A.F.Ioffe Physical-Technical Institute, 194021 St. Petersburg,
Russia}

\author{B.~Spivak}

\address{Physics Department, University of Washington, Seattle, WA 98195, 
USA}

\maketitle

\begin{abstract}
We consider chirality related effects in optical, photogalvanic and
electron-transport properties of carbon nanotubes. We
 show that these  properties of chiral
nanotubes are determined by terms in the electron effective Hamiltonian
describing the coupling between the electron wavevector along the tube
principal axis and the orbital momentum around the tube circumference. We
develop a theory of photogalvanic 
effects and a theory of {\it dc} electric current, which is linear in the
magnetic
field and 
quadratic in the bias
voltage. Moreover, we present analytic
estimations for the natural
circular
dichroism and magneto-spatial effect in the light absorption.
\end{abstract}

\pacs{PACS: 05.20-y, 82.20-w}

\section{Introduction}
\label{sec:1}
Since the discovery of carbon nanotubes \cite{discovery} the
physical properties of these nanostructures have attracted a lot of
attention \cite{general}. Usually carbon nanotubes (CNs) are visualized
as a layer of graphene sheet rolled-up into a cylinder.
 Depending of the
way of the rolling up the cylinder can be chiral or non-chiral.

In media with 
 chiral symmetry it is impossible to distinguish between 
polar and axial
vectors.
 This leads to
the existence of 
the natural optical activity \cite{tasaki}, including the optical
rotatory power and circular dichroism in the absence of magnetic
fields, and any other effect of chirality where a polar vector and an
axial vector (or a pseudovector) are interrelated by some phenomenological
equation. The circular photogalvanic effect (CPGE) is among them.
In this effect a {\it dc} current induced by an electro-magnetic wave of
the complex amplitude {\bf E} is proportional to components of the
axial vector $i ({\bf E} \times {\bf E}^*)$ and thus depends on the
sign of the circular polarization of light. The CPGE was predicted in
\cite{cpge,cpge1} and then studied in bulk gyrotropic crystals
\cite{sturman,book} and semiconductor quantum-well structures
\cite{qw,qw1,qw2}.

Theoretically, achiral magnetic properties and optical absorption
spectra of CNs were investigated by Ajiki and Ando in the
effective-mass approximation \cite{ando,ando1,ando2} (see also
\cite{kane}). The theory of the optical activity of CNs has been
considered
by Tasaki et al. \cite{tasaki} in the framework of the microscopic
tight-binding model.

 In the present paper we consider effects of
chirality in CNs by extending the effective-mass theory
\cite{ando,ando1,ando2}. We show that these effects appear due to terms
in the electron Hamiltonian describing the coupling between the orbital
momentum around the circumference of a chiral CN and the linear
electron momentum along the tube. This is quite different from the case
of bulk materials and semiconductor quantum wells where these effects
are attributed to spin-orbit terms $\sigma_{\alpha} k_{\beta}$ in the
electron effective Hamiltonian ($\sigma_{\alpha}$ and ${\bf k}$ are the
Pauli spin matrices and the electron wave vector; $\alpha, \beta$ are the
cartesian coordinates) \cite{cpge,qw2} (see also \cite{tellurium}).  

The paper is organized as follows. In Sec.\ \ref{sec:2} we briefly
review properties of the single particle electron spectrum of CNs and
derive an expression for the matrix element of the electron-photon
interaction. In Secs.\ \ref{sec:3} and \ref{sec:3a} we present the theory of
photogalvanic effects in chiral CNs. We show that the circularly
polarized light generates a {\it dc} current through CN and the
current direction depends on the sign of polarization, i.e. on the photon
spirality. In the presence of an external magnetic field $B$ parallel
to the nanotube principal axis, the linearly polarized light generates
a {\it dc} current whose direction changes upon the inversion of $B$.
In Sec.\ \ref{sec:3b} we develop a theory of the magneto-chiral anisotropy
in the {\it dc} charge transport through a CN, which is the existence of
a {\it dc} current proportional to $B$ and quadratic in the applied
electric
field $E$. In Sec.\ \ref{sec:4} we give analytical estimates for the 
optical
anisotropy and optical activity of CN. In Sec.\ \ref{sec:5} we discuss
limitations of our approach and mention other effects which are similar to
those considered in this article.

\section{Electron band structure and electron-photon matrix elements in
carbon nanotubes.} 
\label{sec:2}

Usually carbon nanotubes (CNs) are visualized
as a conformal mapping of a graphene sheet onto a cylindrical surface
where one of the two-dimensional (2D) Bravais lattice vectors,
{\bf L}, maps to the cylinder circumference \cite{saito,saito1,white}.
The structures specified by the vectors {\bf L} directed along one of
the two-fold rotation axes $u_2$ or $u'_2$ are called respectively
armchair and zigzag CNs, where the axis $u_2$ is parallel and the axis
$u'_2$ is perpendicular at least to one side of the 2D lattice hexagon.
These particular tubes are achiral. Except for them all others are
chiral with their principal axis $z$ being the screw axis.

Following \cite{ando,ando1,ando2} we write the circumferential vector as
\begin{equation}
{\bf L} = n_a {\bf a} + n_b {\bf b}\:,
\end{equation}
where $n_a, n_b$ are integers and ${\bf a}, {\bf b}$ are the 2D basis
vectors with the angle 120$^{\circ}$ between them. Choosing the
coordinate system $x_1, x_2$ in such a way that $x_1 \parallel {\bf
a}$, $x_2 \perp {\bf a}$ we have for the components of {\bf a} and {\bf b}
\begin{equation}
{\bf a} = a (1,0)\:,\:{\bf b} = a \left(- \frac12, \frac{\sqrt{3}}{2}
\right)\:, 
\end{equation}
where the lattice constant $a$ is equal to $\sqrt{3}$ times the
interatomic distance $d = 1.44$ \AA~\cite{prl01}. The electron
effective Hamiltonian for a graphene sheet,
\begin{equation} \label{hamilt}
H= \left( \begin{array}{ccc}
0 & \mbox{}\hspace{2 mm}\mbox{}&h^{*} \\
h & \mbox{}\hspace{2 mm}\mbox{}&0
\end{array} \right)\:,
\end{equation}
is expanded in the vicinity of the points  
\begin{equation}
{\bf K} = \frac{4 \pi}{3 a} (-1,0)\:,\:{\bf K}' = \frac{4 \pi}{3 a}
(1,0) 
\end{equation}
at the corners of the 2D Brillouin zone shown in Fig.~1. In the
following we define {\bf k} and ${\bf k}'$ as wavevectors referred
respectively to the points ${\bf K}$ and ${\bf K}'$ and assume the
products $ka, k'a\ll 1$ to be small. Then, in the second order in $ka$ or
$k'a$, the matrix element $h$ in Eq.~(\ref{hamilt}) is given by (see
\cite{ando2}) 
\begin{equation} \label{hk}
h ({\bf k}, {\bf K}) = \gamma e^{- i \theta} \left[ k_{\perp} - i
k_z + \frac{a}{4 \sqrt{3}} e^{3 i \theta} \left( k_{\perp} +
i k_z \right)^2 \right] 
\end{equation}
near the {\bf K} point and
\begin{equation} \label{hk'}
h ({\bf k}', {\bf K}') = \gamma e^{i \theta} \left[ - k'_{\perp} - i
k'_z + \frac{a}{4 \sqrt{3}} e^{-3 i \theta} \left( k'_{\perp} -
i k'_z \right)^2 \right] 
\end{equation}
near the ${\bf K}'$ point. Here 
$\gamma = \frac{\sqrt{3}}{2} \gamma_{0} a$,
$\gamma_0$ ($\approx 3$ eV \cite{ando,prl01}) is the transfer integral
between neighboring $\pi$ orbitals, $\theta$ is the angle between the
vector {\bf L} and the basis vector {\bf a}. The subscripts $z, \perp$
indicate components of a vector referred to the axes lying in the
graphene plane and related to the vector {\bf L} so as $z
\perp {\bf L}$ and $k_z \perp {\bf L}$, $k_{\perp} \parallel {\bf
L}$. 

In the same approximation the energy spectrum near the {\bf K}
point is given by 
\begin{equation} \label{spectrum}
E_{c,v}({\bf k}, {\bf K}) = \pm \vert h \vert \approx \pm \gamma 
\left\{ |{\bf k}| + \frac{a}{4 \sqrt{3} |{\bf k}|}
\left[ \left(k_{\perp}^3 - 3 k_{\perp} k_z^2\right) \cos {3 \theta } + 
\left( k_z^3 - 3 k_{\perp}^2 k_z \right) \sin {3 \theta } \right]
\right\} \:,
\end{equation}
where $|{\bf k}|$ = $\sqrt{k_{\perp}^2 + k_z^2}$, the upper and lower
signs represent the conduction (subscript $c$) and valence (subscript
$v$) bands respectively. The similar spectrum near the ${\bf K}'$ point
is obtained by changing $k_{\perp} \rightarrow - k'_{\perp}$, $k_z
\rightarrow k'_z$, $\theta \rightarrow - \theta$ in agreement with the
time inversion symmetry requirement $E_{c,v}({\bf k}, {\bf K}')$ $=
E_{c,v}(- {\bf k}, {\bf K})$.  

In a CN specified by the vector {\bf L} the electron wave function
satisfies the cyclic boundary condition $\Psi({\bf r}) = \Psi({\bf
r+L})$. This enables one to find the allowed discrete values of
$k_{\perp}$ as 
\begin{equation} \label{allowed}
k_{\perp} = \frac{2 \pi}{L} \left( n - \frac{\nu}{3} \right)\:; \:
k'_{\perp} = \frac{2\pi}{L} \left( n + \frac{\nu}{3} \right), 
\end{equation}
where $n$ is an integer $0, \pm 1, \pm 2...$ characterizing the angular
momentum component of an electron, $L=|{\bf L}| = a \sqrt{n_{a}^{2} +
n_{b}^{2}-n_{a}n_{b}}$, and $\nu$ equals to one of three integers: $0,
\pm 1$ determined by the presentation of the sum $n_{a}+n_{b}$ as
$3N+\nu$ with integer $N$. The dispersion in the conduction and
valence
subbands is obtained by substituting Eq.~(\ref{allowed}) into
Eqs.~(\ref{hk},\ref{hk'}) or Eq.~(\ref{spectrum}).
In the following we focus on the nanotubes characterized
by finite band gap and assume $\nu \neq 0$.
According to Eq.~(\ref{spectrum}), for small values of $k_z$ satisfying
the condition $|k_z| \ll |k_{\perp}|$ in the $K$ valley and similar
condition in the $K'$ valley, the electron spectrum has a parabolic
form with terms linear in $k_z$
\begin{eqnarray} \label{021}
E_{c,v}(n, k_z; K) &=& \pm \left( \frac{\Delta_n}{2} +
\frac{\hbar^2 k_z^2}{2 m_n} + \beta_n k_z \right)\:, \\
E_{c,v}(n, k'_z; K') &=& \pm \left( \frac{\Delta'_n}{2} +
\frac{\hbar^2 k_z^{'2} }{2 m'_n} + \beta'_n k'_z \right)\:, \nonumber 
\end{eqnarray}  
where
\begin{equation} \label{delta}
\Delta_n =  2 \gamma |k_{\perp}|\:,\:
\Delta'_{n} = 2 \gamma |k^{'}_{\perp}|\:,
\end{equation}
\begin{equation} \label{mass}
m_n = \frac{\hbar^2 |k_{\perp}|}{\gamma}\:,\:
m'_{n} = \frac{\hbar^2 |k'_{\perp}|}{\gamma}\:,
\end{equation}
\begin{equation} \label{beta}
\beta_n = - \frac{\sqrt{3}}{4} \gamma a |k_{\perp}| \sin{3 \theta}\:,\:
\beta'_{n} = \frac{\sqrt{3}}{4} \gamma a |k'_{\perp}| \sin{3 \theta}\:,
\end{equation}
and $k_{\perp}, k'_{\perp}$ are defined in Eq.~(\ref{allowed}). Note
that the identity $E_{c,v}(n, k_z; K')$ $= E_{c,v}(-n, - k_z; K)$
follows directly from the time inversion symmetry.

In the presence of an external magnetic field ${\bf B}$, the electron
energy is modified just by changing $k_{\perp}, k'_{\perp}$ from
Eq.~(\ref{allowed}) into
\begin{equation} \label{kmagn}
k_{\perp} = \frac{2 \pi}{L} \left( n - \frac{\nu}{3}
+\frac{\Phi}{\Phi_{0}} \right)\:; \:
k'_{\perp} = \frac{2\pi}{L} \left( n + \frac{\nu}{3}
+\frac{\Phi}{\Phi_{0}} \right)\:,
\end{equation}
where $\Phi$ is the magnetic flux passing through the cross section of
a CN, $\Phi = B_z L^2 / (4 \pi)$, and $\Phi_0$ is the magnetic flux
quantum, $c h /e$. Now the consequence of the time inversion symmetry
takes the form $
E_{c,v}(n, k_z; \Phi ; K')$ $= E_{c,v}(-n, - k_z; - \Phi ; K)\:.$

Chirality (or spirality) of a nanotube manifests itself in a particular
coupling between the angular momentum as described by $n$ and the
directed translational motion as described by $k_z$: due to the
terms linear-$k_z$ in Eq.~(\ref{021}) or, in general, due to
odd-in-$k_z$ terms in Eq.~(\ref{spectrum}) the energy has a
contribution which depends both on the $sign$ of $k_z$ and the $sign$
of $n$.

 It is interesting to analyze how this particular coupling
disappears for zigzag and armchair tubes which are achiral from the
symmetry point of view. In zigzag tubes, the angle $\theta$ between the
circumferential vector {\bf L} and the vector {\bf a} is an integer
multiple of 60$^{\circ}$, $\sin{3 \theta}$ is zero and odd-in-$k_{z}$
terms
in Eqs.~(\ref{spectrum},\ref{021}) vanish. In armchair tubes, the angle
$\theta$ equals to $\theta = 30^{\circ} + N \cdot 60^{\circ}$ with $N$
integer leading to one of the following three relations between $n_a$
and $n_b$: $n_a = 2 n_b$ or $n_b = 2 n_a$ or $n_b = - n_a$. This means
that the sum $n_a + n_b$ is an integer multiple of 3 and the parameter
$\nu$ is zero. As a result values of $|k_{\perp}|$ become independent
of the sign of $n$ and the coupling between signs of $n$ and $k_z$ in
the terms odd-in-$k_z$ terms. This follows also from the
symmetry
considerations and is an important check point. In the following
sections we will show that the odd-in-$k_z$ terms in the electron
energy spectrum govern the chirality effects in CNs.

Let us now turn to the calculation of the matrix element, $V_{fi}$, for
the
electron optical transition between the initial state $i$ and the final
state $f$.
The electron
envelope functions can be written in the form
\begin{equation} \label{envelope}
\psi_{c,v} (z;n,k_z) = \frac{e^{i n \varphi}}{\sqrt{2 \pi}}\:
\frac{ e^{i k_z z} }{\sqrt{L_{CN}}}\: \hat{C}_{c,v}(n,k_z)\:,
\end{equation}
where $L_{CN}$ is the nanotube length and $\varphi$ is the azimuth
angle. The two-component columns $\hat{C}_{c,v}$ are eigenvectors of
the 2$\times$2 matrix Hamiltonian (\ref{hamilt}),  given by
\begin{equation}
\hat{C}_{c} = \frac{1}{\sqrt{2}} \left[ \begin{array}{c} h^* / |h|
\\ 1 \end{array} \right]\:,\:\hat{C}_{v} = \frac{1}{\sqrt{2}}
\left[ \begin{array}{c} 1 \\ - h / |h| \end{array}\right]
\end{equation}
with $h$ defined by Eq.~(\ref{hk}) for the $K$ valley and by
Eq.~(\ref{hk'}) for the $K'$ valley where $K, K'$ are the
$z$-components of the vectors ${\bf K}, {\bf K}'$.

 We remind that the electron interaction with the
electromagnetic field is described by the perturbation $(e/c) \{
\hat{{\bf v}} {\bf A} \}_s$, where {\bf A} is the vector potential of
the field, $- e$ is the electron charge, $\hat{{\bf v}}$ is the
electron velocity operator $\hbar^{-1} d H / d k_z$ and the symbol
$\{...\}_s$ means a symmetrized product of operators. If $|k_z| \ll
|k_{\perp}|$ then one can neglect the second-order terms in the
expansion Eq.(5), the symmetrization symbol can be omitted and the
scalar product of $\hat{{\bf v}}$ and the light unit polarization
vector {\bf e} has the form 
\begin{equation} \label{ve}
\hat{{\bf v}} \cdot {\bf e} = \frac{i \gamma}{2 \hbar} \left[
\begin{array}{cc} 0& e^{i \theta} f_{12} \\
e^{-i \theta} f_{21}&0 \end{array} \right]\:.
\end{equation}
Here
\[
f_{12} = e^{i \varphi} e_- - e^{-i \varphi} e_+ + e_z\:,\:
f_{21} = e^{i \varphi} e_- - e^{-i \varphi} e_+ - e_z\:,
\]
$e_{\pm} = e_x \pm i e_y$ and $x,y$ are rectangular axes perpendicular
to the principal axis $z$ of the tube. The selection rules for matrix
elements
of the operator taken between the envelope functions Eq.(16)
are in agreement with the conservation law for $z$-components of the
angular momenta: $n_f =$ $n_i + n_{phot}$, where $n_{phot} = \pm 1$ for
the circular polarization $\sigma_{\pm}$ of a photon propagating along the
$z$ axis and $n_{phot} = 0$ for the light linearly polarized along $z$.
In particular, the transitions $(c, 0, K) \rightarrow (c, 1, K)$, $(c,
0, K') \rightarrow (c, - 1, K')$ occur respectively under the
$\sigma_+$ and $\sigma_-$ photoexcitation. The squared moduli of the
corresponding matrix elements are given by 
\begin{equation} \label{vele}
|(\hat{{\bf v}} \cdot {\bf e})_{fi}|^2 = \frac18 \left[
\frac{\gamma}{\hbar} k_z \left( \frac{1}{k_{\perp, 0}} -
\frac{1}{k_{\perp, 1}} \right)\right]^2 = \frac{1}{32} \left(
\frac{\gamma}{\hbar}\: \frac{k_z}{k_{\perp, 0}} \right)^2 \:,
\end{equation}
where we took into account that $k_{\perp, 1} = 2 k_{\perp, 0}$. The
transitions under consideration are forbidden at the ${\bf K}$
and ${\bf K}'$ points but become allowed for $k_z, k'_z \neq 0$. The
squared moduli of the optical matrix element is given by
\begin{equation}
|V_{fi}|^2 = \left( \frac{e}{c} A \right)^2 |(\hat{{\bf v}} \cdot {\bf
e})_{fi}|^2 = \frac{2 \pi e^2 I}{\omega^2 c n_{\omega}}
\frac{1}{32} \left( \frac{\gamma}{\hbar}\: \frac{k_z}{k_{\perp, 0}}
\right)^2\:,
\end{equation}
where $A = |{\bf A}|$, $I$ is the light intensity and $n_{\omega}$ is
the refractive index of the medium. In the following we consider a
free-standing nanotube and assume $n_{\omega} = 1$.

It should be stressed that the linear-optics approximation is valid
as long as $\hbar^{-1} |V_{fi}|\ll (2 \tau_p)^{-1}$,
otherwise
one has to take into account the saturation of the absorption.
 Another important point is that the
depolarization effect \cite{tasaki,ando1} can be neglected provided 
\begin{equation} \label{depol}
\left\vert \frac{4 \pi \sigma_{\perp}}{L \omega} \right\vert \ll 1\:,
\end{equation}
where $\sigma_{\perp}$ is the conductivity at the frequency $\omega$ 
defined as 
\[
\hbar \omega W = 2 \sigma_{\perp} \left( \frac{\omega}{c} A
\right)^2
\]
with $W$ being the optical transition rate. 

\section{Circular photogalvanic effect} \label{sec:3}
Physically, the CPGE can be considered as a transformation of the
photon angular momenta into a translational motion of free charge
carriers. It is an electron analog of mechanical systems which
transmit rotatory  motion to linear one 
like a
screw tread or a plane with a propeller. Phenomenologically, in
the
case of chiral CNs it is described by 
\begin{equation}
\label{eq1}
j_{CPGE,z} = \Gamma\:i ({\bf E} \times {\bf E}^* )_{z} \:,
\end{equation}
where $ j_{CPGE,z}$ is the {\it dc} photocurrent, $\Gamma$ is a real
coefficient, {\bf E} is the complex amplitude of the electric field of
the electromagnetic wave and, for the transverse wave,
\begin{equation} \label{pcirc}
i\: ({\bf E} \times {\bf E}^* ) = P_{circ}\:E_0^2\: \hat{o}
\end{equation}
with $E_0, P_{circ}$, $\hat{o}$ being respectively the 
electric field amplitude $|{\bf E}|$, the degree of the circular
polarization of light and the unit vector pointing in the direction of
light
propagation. The photocurrent (\ref{eq1}) reverses its direction under
inversion of the light circular polarization and vanishes for
linearly-polarized excitation. 

Usually the microscopic theory of natural optical activity in bulk
semiconductors \cite{tellurium,cds} as well as the theory of CPGE
\cite{cpge,qw2} is 
based on allowance of spin-dependent linear in ${\bbox k}$ terms 
$\beta_{lm}\sigma_l k_m$ in the electron effective Hamiltonian, where
${\bbox k}$ is the electron wavevector and $\sigma_l$ are the Pauli spin
matrices. The real coefficients $\beta_{lm}$ form a pseudotensor
subjected to the same symmetry restriction as the pseudotensors
describing the optical activity and CPGE. In CNs the spin-orbit
interaction is negligible
and the similar role is played by the coupling between the orbital angular
momentum $n$ and the wave vector $k_{z}$ as described by
Eqs.~(\ref{spectrum},\ref{021}). 

The transfer of photon angular momenta into an electric
current along the principal axis of a chiral CN can be
described by the standard equation for the current 
\begin{equation} \label{currentcv}
j_{CPGE,z} = - e \sum_{n, k_{z}, s} v_c(n, k_{z}, s) f_{c} (n,
k_{z}, s)
+ e \sum_{n, k_{z}, s} v_{h}(n, k_{z}, s ) f_h (n, k_{z}, s )\:,
\end{equation}
where $e$ is the elementary charge ($e > 0$), the index $s$ labels
the valleys $K$ and $K'$, $v_c(n, k_{z}, s )=\hbar^{-1} (dE_c(n, k_{z},
s) / d k_{z})$  is the group velocity and $f_c(n, k_{z}, s
)$ is the nonequilibrium steady-state distribution function for
electrons in the conduction band. The similar quantities for holes in
the valence band $v$ have the subscript $h$. Note that they are related
with the corresponding quantities in the electron representation of the
valence states by $v_h(n, k_{z}, K) = - v_v(- n, - k_{z}, K')$, $f_h (n,
k_{z}, K) = 1 - f_v (- n, - k_{z}, K')$.

In the following we will consider the case where elastic scattering
processes are more effective than the inelastic ones. 
Then one can apply the standard procedure of solving the kinetic equation
and decompose the distribution function $f_{m}$ $(m = c, h)$ into the 
contributions 
\begin{equation} \label{ktilde}
f^{\pm}_{m} (n, k_z, s) = \frac12 \: [f_{m} (n, k_{z}, s) \pm
f_{m} (n, \tilde{k}_{z}, s)]
\end{equation}
that are even and odd with respect to the change of $k_{z}$ by
$\tilde{k}_z$ where $\tilde{k}_z$ belongs to the same valley and
satisfies the equation $E_m(n, \tilde{k}_{z}, s)$ $= E_m(n, k_{z},
s)$. Since the even contribution in Eq.~(\ref{ktilde}) is in fact a
function of the electron energy, it nullifies the elastic scattering
integral and also makes no contribution to the current. Introducing
the momentum relaxation times $\tau_p^{(f,i)}$ we can present the
photocurrent as
\begin{equation} \label{currentfi}
j = - e \sum_{n_f, n_i, k_{z}, s} \left[ v_c(n_f, k_{z}, s
) \tau_p^{(f)}
- v_c(n_i, k_{z}, s ) \tau_p^{(i)} \right] W_{cc}(n_f , n_i, k_{z},
s) 
\: \nonumber
\end{equation}
for optical transitions $(c, n_i, s)$ $\rightarrow (c, n_f, s)$
between the conduction subbands and in a similar way for transitions
between the valence subbands or for interband transitions, where $n_f,
n_i$ are the angular momentum components in the final and initial
states. The transition rate is given by
\[
W_{cc}(n_f , n_i, k_{z}, s) = \frac{2 \pi}{\hbar} \vert V_{fi} \vert^2\:
f_c^0[E_c(n_i, k_{z}, s)]\: \delta [E_c(n_f, k_{z}, s) - E_c(n_i,
k_{z}, s)
- \hbar \omega ] \:.
\]
Here $f_c^0$ is the equilibrium distribution function and we assume
that in equilibrium the upper subband $(c, n_f, s)$ is unoccupied. 

In this section we take a chiral CN with the parameter $\nu = 1$,
assume that the tube is $n$-doped and consider the intersubband
photoexcitation of electrons from the lowest conduction subbands $(c,
n=0, s)$ to the first higher subbands $(c, n = 1, K)$ and $(c, n=-1, 
K')$ (see Fig.~2). In order to obtain compact analytical results we
consider near-edge optical transitions with the light frequency
$\omega$ satisfying the condition 
\begin{equation}
|\hbar \omega - \Delta_{10}| \ll \Delta_{10} \:,
\end{equation}
where $\Delta_{10}$ is the energy separation between the subbands 
$(c, 0, K)$ and $(c, 1, K)$ given by, see
Eqs.~(\ref{allowed},\ref{delta}),
\[
\Delta_{10} = \frac12 (\Delta_1 - \Delta_0) = \frac{2 \pi \gamma}{3 L}.
\]
In this case $|k_{z}| \ll |k_{\perp}|$ and one can use the approximate
equations (\ref{021})--(\ref{beta}). Rewriting Eq.~(\ref{021}) in the form
\begin{equation}
E_{c}(n, k_{z}; K) = \frac{\Delta_n}{2} - \frac{m_n \beta_n^2}{2 \hbar^2}
+ \frac{\hbar^2 ( k_{z} + \kappa_n)^2}{2 m_n} 
\end{equation}
with $\kappa_{n} = m_n \beta_n / \hbar^2$ we conclude that $\tilde{k_{z}}$
entering Eq.~(\ref{ktilde}) equals to $- k_z - 2 \kappa_n$. We ignore
the small frequency region where the terms linear-in-$ k_z$ exceed
or are comparable with the quadratic terms in-$k_z$ in
Eq.~(\ref{021})
and assume the ratios $\ae_n(k_z)$ $= \beta_n k_z / (\hbar^2 k_z^2
/ 2 m_n)$ for $n=0,1$ to be small which is valid if 
\[
|k_z| \gg 2 m_n \beta_n / \hbar^2\:\mbox{or}\: |\hbar \omega -
\Delta_{10}| \gg 2 m_n \beta_n^2 / 
\hbar^2 . 
\]

Taking into account Eq.~(\ref{vele}) and retaining the first-order
terms in $\ae_n$ we can write the photocurrent as a sum of four
contributions 
\begin{equation} \label{currentf}
j_{CPGE} = e W_{1,0} (l_v + l_m + l_{\tau} + l_f)\:.
\end{equation}
Here $W_{1,0}$ is the transition probability rate per unit length
calculated neglected the terms linear-$k$:
\begin{equation}
W_{1,0} = \frac{2 \pi}{\hbar} \frac{2 \pi e^2 I}{\omega^2 c n_{\omega}}
\frac{1}{32} \left( \frac{\gamma}{\hbar}\: \frac{k_z}{k_{\perp, 0}}
\right)^2 f_c^0(E_0)\: g_{10} (\Delta_{10} - \hbar \omega)\:,
\end{equation}
$I$ is the light intensity in units $energy \cdot length^{-1} \cdot
time^{-1}$, the
one-dimensional reduced density of states equals to 
\[
g_{10} (E) = 2 \times \sum_{k_z} \delta(- \frac{\hbar^2
k_z^2}{2 \mu_{10}} - E) = \frac{1}{\pi} \left( \frac{2
\mu_{10}}{\hbar^2 E} \right)^{1/2} \: \theta (- E),
\]
the factor of two makes allowance for the spin degeneracy, $\theta(x)$
is the step function equal to 0 if $x<0$ and 1 if $x > 0$, $-
\mu_{10}^{-1}$ is the inverse reduced effective mass $m_1^{-1} - 
m_0^{-1}$, (since $m_1 = 2 m_0 > m_0$ a value of $\mu_{10} = 2 m_0$ is
positive and $g_{10} (E)$ is defined for negative values of $E$), and
$f_c^0(E_0)$ is the value of the equilibrium distribution function at the
energy $E_0 =$ $(\hbar^2 k_z^2 / 2 m_0)$ $=( \mu_{10} / m_0)
(\Delta_{10} - \hbar \omega)$ $= 2 (\Delta_{10} - \hbar \omega)$. 
The lengths $l_v, l_m, l_{\tau}, l_f$ in Eq.~(\ref{currentf}) are
related to the photoexcitation asymmetry arising due to the
$k_z$-dependence of the velocity and density of states ($l_v$), of the
squared matrix element ($l_m$), of the momentum relaxation time
($l_{\tau}$) and of the equilibrium distribution function ($l_f$).
For the optical transitions under consideration the straightforward
derivation results in 
\begin{equation}
l_v = \frac{3 \beta_0}{\hbar} (\tau_p^{(1)} - \tau_p^{(0)}) \:,\: 
l_m = \frac{2 \beta_0}{\hbar} (\tau_p^{(1)} - 2 \tau_p^{(0)}) \:,
\end{equation}
\[
l_{\tau} = \frac{6 \beta_0}{\hbar} \left( \tau_p^{(1)} 
\frac{d \ln{\tau_p^{(1)}}}{d \ln{E_1}}  - \tau_p^{(0)}
\frac{d \ln{\tau_p^{(0)}}}{d \ln{E_0}} \right)\:,\:
l_f = \frac{3 \beta_0}{\hbar} (2 \tau_p^{(0)} - \tau_p^{(1)})
\frac{\Delta_{10} - \hbar \omega}{k_B T} [1 - f_0(E_0)]\:.
\]
Obviously, the momentum relaxation time $\tau_p^{(1)}$ is shorter
than $\tau_p^{(0)}$ because a photoelectron excited to the subband
($(c, n=1$) can be readily scattered to the subband ($(c, n=0$).

The circular photocurrent can be estimated as
\begin{equation} \label{cpgefin}
j_{CPGE} \sim e W_{1,0}\: \frac{\beta_0}{\hbar} \tau_p \: P_{circ}
\end{equation}
and for intensities at the absorption saturation
\[
j_{CPGE} \sim \frac{2 \pi \sin{3 \theta}}{\sqrt{3}} \frac{e}{\tau_p} 
\frac{a}{L} \left( \frac{\Delta_{10}}{\Delta_{10} - \hbar \omega}
\right)^{1/2}\:.
\]
Taking $\gamma=$ 6.5 eV$\cdot$\AA, $a$ = 2.5~\AA,
$\nu = 1$, $L =$ 135~\AA, $\sin{3 \theta} =$ 0.7 we have
$\Delta_{10} = 0.1$, $m_{0}=0.017 m$, where $m$ is the bare 
electron mass, $\beta_0/ \hbar =$ 1.1$\times$10$^{6}$ cm/s.
Then for $\tau_p =$ 10$^{-11}$ s, 
$\Delta_{10} - \hbar \omega = 0.1 \:\Delta_{10}$, 
and $f_c^0(E_0) \sim 1$ we obtain $j_{CPGE} \sim 10^{-9}$ A. 
While making this estimation we assumed the difference $\tau_p^{(0)} -
\tau_p^{(1)}$ and the electron-electron scattering time $\tau_{ee}$ to
be comparable with $\tau_p^{(0)}$. It is worth to mention
that a value of the photocurrent in recently discovered
crystals of CNs \cite{schlittler} will be significantly enhanced
compared to a single CN.

We conclude the section by comparing the CPGE described by
Eq.~(\ref{cpgefin}) with the photon drag (PD) effect which exists
in crystals of arbitrary symmetry and is independent on the sign
of the circular polarization. The PD current is estimated as
\begin{equation}
j_{PD} \sim e W_{1,0}\: \frac{\hbar q}{m_0} \tau_p \:,
\end{equation}
where $q$ is the photon wave vector (in vacuum $q = \omega /c$).
Thus, for $\hbar \omega =$ 0.1 meV we have
$j_{CPGE} /j_{PD} \sim \beta_{0} m_{0}/ (\hbar^2 q) \approx 3$.

In this section we presented both the detailed calculation and
estimation by an order of magnitude for the CPGE. In the next chapters
we restrict ourselves only to analytical estimations of other effects
while their detailed consideration can be given elsewhere.

\section{Magneto-induced linear photogalvanic effect.} 
\label{sec:3a} 
In addition to the circular PGE, in noncentrosymmetric media a
photocurrent of another kind can be induced by the electro-magnetic
wave. This is called the linear photogalvanic effect (LPGE) and described
by a third-rank tensor, $\chi_{ijk}$, symmetrical with respect to the
interchange of the indices $j$ and $k$. Thus, for the LPGE one has
\cite{book} 
\begin{equation} \label{lpge}
j_{LPGE,i}= \chi_{ijl}\:I\: (e_{j}e^{*}_{l}+e^{*}_{j}e_{l})/2 \:.
\end{equation}
Ideal CNs are unpolar with their principal axis, $z$, being two-sided which
forbids the photocurrent (\ref{lpge}). In B$_x$C$_y$N$_z$ nanotubes
(BNs) the symmetry is reduced, the axis $z$ is polar and the LPGE
becomes allowed  \cite{tomanek}. We show here that the
linear photocurrent can be induced in an ideal CN in the presence of an
external magnetic field ${\bf B} \parallel z$. Phenomenologically the
magneto-induced LPGE can be written as
\begin{equation} \label{lpgecn}
j_{M\mbox{-}LPGE, z} = I B_z \:[\Lambda_{\parallel} |e_z|^2 +
\Lambda_{\perp} (|e_x|^2 + |e_y|^2)] 
\end{equation}
and determined by two linearly independent coefficients
$\Lambda_{\parallel}, \Lambda_{\perp}$.

For the interband transitions $(v, 0) \rightarrow (c, 0)$ excited in an
undoped CN by the linearly polarized light, ${\bf e} \parallel z$, one
has the following estimation
\begin{equation} \label{jlpge}
j_{M\mbox{-}LPGE,z} \sim e W_{0,0}\: \frac{\beta_0}{\hbar} \tau_p \:
\nu\: \frac{\Phi}{\Phi_0}\:.
\end{equation}
The estimation follows if we take into account the quadratic term in the
expansion of $h$ in powers of $ka$. Due to this term the squared
matrix element of the optical transition has an odd contribution, the
ratio of the odd contribution to the main even contribution being
proportional to $a k_z \sin{3 \theta}$. In the magnetic field the band
gaps $\Delta_{00}(K), \Delta_{00}(K')$ differ due to the difference
of $|k_{\perp}|$ and $|k'_{\perp}|$ at $n=0$ (See Eq.~(\ref{kmagn})
and Fig.3 where the scheme of the optical transitions is shown. The
different thickness of lines in Fig.3, which correspond to the
transitions of electrons with different momenta indicate that
they probabilities are different). This
leads to the relative difference in the transition rates for the $K$ and
$K'$ valleys proportional to
\[
\frac{\Delta_{00}}{\hbar \omega -
\Delta_{00}}\:\nu\:\frac{\Phi}{\Phi_0} \:. 
\]
Thus the linear magneto-photocurrent induced by the light polarized
along the principal axis $z$ can be estimated as
\[
j_{M-LPGE,z} \sim e W_{0,0}\: a k_z \sin{3 \theta}\: \frac{\hbar k_z}{m_0}
\:\frac{\Phi}{\Phi_0} \:\frac{\Delta_{00}}{\hbar \omega -
\Delta_{00}}\: \tau_p\:.
\]
Since
\[
k_z \frac{\hbar k_z}{m_0} \sim \frac{\hbar \omega - \Delta_{00}}{\hbar} 
\]
and $\Delta_{00} \sim \gamma k_{\perp, 0}$, $\beta_0 \sim a \gamma
\sin{3 \theta}$ we obtain 
\[
j_{M-LPGE,z} \sim e W_{0,0}\:\frac{\Delta_{00} a}{\hbar} \tau_p \sin{3
\theta} \:\frac{\Phi}{\Phi_0} 
\]
and finally arrive at Eq.~(\ref{jlpge}). The same order of magnitude
for $j_{M-LPGE,z}$ is obtained if we ignore the quadratic terms in $h({\bf
k})$ but include the cubic-$k_z$ terms in the energy dispersion
Eq.~(\ref{spectrum}). Above we assumed $\Phi$ to be much smaller than
$\Phi_0$. For large values of $\Phi$ the linear magneto-photocurrent is
a periodic function of the ratio $\Phi/\Phi_{0}$. 
 While preparing the manuscript we learned about a similar work
\cite{kibis} on 
the magneto-induced linear photovoltaic effect.

\section{Magneto-induced \lowercase{{\it dc}} electric current
\protect\\ quadratic in the electric field} \label{sec:3b}

Eq.(35) for  magneto-induced
linear PGE holds at high frequencies $\omega > \tau_p^{-1}$. This
effect, however, does not vanish even in
the case $\omega=0$: the chiral symmetry
allows
existence of a current  which is  quadratic in the
electric field and linear in the external magnetic field.
In the case of bulk metals this effect has been observed in 
\cite{rikkenNL}.

In the case of chiral CN a general expression for {\it dc} current
has the form 
\begin{equation}
j_z = \sigma E_z + \Lambda E_z^2 B_z\:,
\end{equation}
where $\sigma=e^2 n \tau_p/ m_0$ is the linear Drude conductivity,
$\tau_p$ is the momentum relaxation time and $m_0$ is the effective
mass in the lowest conduction subband, and we assume the Fermi level
lies below the bottom of the subbands $(c, \pm 1)$. Note that the
magneto-chiral coefficient $\Lambda$ is nonzero only in chiral CNs.

 Qualitatively, the nature of the magneto-chiral
correction can be understood as follows: the electric field accelerates
electrons (in $n$-doped samples) and creates a nonequilibrium
electron distribution. In this section we assume that the elastic 
electron relaxation rate is much larger than the
inelastic one , which is associated with the electron-phonon scattering. 
Than in the first approximation in the ratio between these rates
the electron distribution function depends only on the electron energy.
 It should be
determined from
the balance of the energy supplied to the electron system by the electric 
field and the energy transfered by electrons to the acoustic phonons.
It is important that the distribution functions which depend only on the
electron
energy  correspond to zero current.
To get a nonzero value of the {\it dc} current we have to take into
account
that in the case 
of chiral CN the electron-phonon inelastic scattering rate depends on the 
direction of the electron momentum.
 As a result, there is
a correction to
the electron distribution function which is
odd-in-$k_{z}$, the value of which is proportional to the ratio between
the
inelastic and elastic relaxation rates.
 In other words, the electron-phonon inelastic processes convert
a currentless distribution of the electron gas into a distribution
characterized by a nonzero electric current, proportional to $E^{2}$.

An antisymmetric part of the probability rate for the electron
inelastic scattering can be found if we use the symmetry considerations
and write, in addition to the 2$\times$2 Hamiltonian (\ref{hamilt}),
the operator of the electron-phonon interaction for a graphen sheet near
the point {\bf K} 
\begin{equation} \label{elphon}
V_{e-phon} = \left( \begin{array}{cc}
\Xi_0\: u & \Xi \:u_+ \\
\Xi \:u_- & \Xi_0 \:u
\end{array} \right)\:.
\end{equation}
Here $u = u_{11} + u_{22}$, $u_{\pm} = u_{11} - u_{22} \pm i u_{12}$,
$u_{lm}$ is the in-plane strain tensor in the axes $x_1, x_2$ and
$\Xi_0, \Xi$ are the deformation potential constants. In the
tight-binding model $\Xi = - \Xi_0/2$. Let $u_{zz}$ be a uniform strain
in a single CN. Then the strain-induced shift of the electron energy in
the subband $(c,0)$ in the $K$ valley is equal to 
\begin{equation}
\varepsilon_u (\theta; K) = \left[ \Xi_0 - \frac{\sqrt{3}}{2} \Xi
\left( \frac{k_{\perp,0}}{k_0} \cos{3\theta} + \frac{k_z}{k_0}
\sin{3\theta} \right) \right] \: u_{22} 
\end{equation}
and $\varepsilon_u (\theta;K')= \varepsilon_u (-\theta;K)$. Here
$k_0 = \sqrt{k_{\perp,0}^2 + k_z^2}$.
It follows from the above equation that the ratio of the part odd-in-$k_z$
 to the part even-in-$k_z$ for the rate of electron scattering by
a phonon can be estimated as
\begin{equation} \label{oddkz}
\eta(k_z) \equiv \frac{W^{(-)}}{W^{(+)}} \sim \frac{k_z}{k_0}
\sin{3\theta} \:.
\end{equation}
At zero magnetic field the asymmetry induced by inelastic processes in
the $k_z$ distribution in the valley $K$ is compensated by the
asymmetry in the $K'$ valley and the quadratic-in-$E_z$ current is zero.
The magnetic field shifts the bottoms of the $K$ and $K'$ valleys
relative to each other (See Eqs. (9,13)), the equilibrium electron
densities
in these
valleys become different leading to a current contribution proportional
to $E_z^2 B_z$.

In contrast to the previous consideration of free-standing CNs we
consider here a CN that lies on a solid surface and is effected by 3D
acoustic phonons of the solid. It allows to uncouple the momentum and
energy conservation laws in electron-phonon scattering processes. Then
the correction to the {\it dc} current proportional to $E^2 B_z$ is
estimated as 
\begin{equation} \label{e2b}
\delta j_z \sim e \langle W^{(+)} \eta(\bar{k}_{z}) \zeta [\tau_p(k_{f,z})
v(k_{f,z}) - \tau_p(k_{i,z}) v(k_{i,z})] \rangle\:,
\end{equation}
where $\delta j_{z}$ in the part of the current proportional to
$E^{2}_{z}$,
\begin{equation}
\zeta \sim \frac{\gamma k_{\perp,0}}{E_F} \frac{\Phi}{\Phi_0} \:,
\end{equation}
is the relative difference of the
electron equilibrium densities in the valleys $K$ and $K'$ induced by
the magnetic field, see Eqs.~(\ref{delta},\ref{kmagn}),
$E_F$ is the Fermi energy referred to the conduction band bottom,
$\tau_p$ is the momentum relaxation time, $v(k_z) = \hbar k_z / m^*$ is
the electron velocity, $k_{f,z}$ and $k_{i,z}$ are the initial and
final values of $k_z$ in the inelastic scattering process, we assume
the difference $k_{f,z} - k_{i,z}$ to be much smaller than $\bar{k}_z =
(k_{f,z} + k_{i,z})/2$; the angle brackets mean the averaging over $k_z$.
A value of $W^{(+)}$ is estimated from the balance of energy as
\begin{equation} \label{balance}
k_B T W^{(+)} = \sigma E_z^2\:, 
\end{equation}
where $k_B$ is the Boltzmann's
constant and we take into account that the typical electron energy
change under scattering is of the order of the thermal energy or,
analytically, 
\[
\hbar^2 \bar{k}_z (k_{i,z} - k_{f,z}) / m^* \sim k_B T\:.
\]
Here we consider the degenerate electron gas and assume $k_B T$ to be
much smaller than the Fermi energy $E_F$. Then Eq.~(\ref{e2b}) can be
reduced to 
\begin{equation} \label{finale2a}
\delta j_z \sim e \sigma E^2 \sin{3 \theta}\: \tau_p
\frac{\gamma}{\hbar E_F} \frac{\Phi}{\Phi_0} 
\end{equation}
or
\begin{equation} \label{finale2b}
\frac{\delta j_z}{j_z} \sim \sin{3 \theta}\: \frac{e E \tau_p }{\hbar}
\frac{\gamma}{E_F} \frac{\Phi}{\Phi_0} \:.
\end{equation}

We would like to mention that in the approximation of purely elastic
scattering $\delta j_{z}$ is zero. Nevertheless
Eqs.~(\ref{finale2a},\ref{finale2b}) is independent of inelastic
scattering rate. This is because we considered the case when the sample
length is larger than the inelastic diffusion length, $\sqrt{D_{p} 
\tau_{in}}$, where $D_{p} \sim v_{F}^{2}\tau_{p}$ 
is the elastic diffusion coefficient and $v_{F}$ is the Fermi velocity,
and we used the equation (\ref{balance}).
In the opposite limit the value of the current is proportional to
$\tau^{-1}_{in}$.

Assuming, for example, that the voltage bias on the sample is
$U \sim 0.01 V$, $E_{F} \sim 0.1 eV$, $\tau_p \gamma / \hbar \sim
L_{CN}$ and, hence, $E_z \tau_p \gamma / \hbar \sim U$,
$\Phi / \Phi_{0} \sim 10^{-2}$ (for $B_z =$ 10 T,
$L = 70\: \AA$), we get the estimate $\delta j_z/j_z \sim 10^{-3}$.
\section{Chirality effects in optical spectroscopy}
\label{sec:4}
In this section we presen estimates for two other effects
of chirality in optical spectroscopy of CNs: the natural circular dichroism
and the magneto-chiral dichroism.

 Let us start with the natural
circular dichroism appearing as a difference in the interband or intersubband
optical transition rates for the $\sigma_+$ and $\sigma_-$ light
polarizations. As mentioned above, in bulk semiconductors the theory of
natural optical activity is based on allowance for coupling between the
electron spin and wavevector. Contrary to the approach
\cite{tellurium},
we ignore the spin-orbit interaction in CNs because it is negligible for
C atoms and take into account the coupling between the angular momentum $n$
and the wavevector $k_z$. We assume the circularly polarized light to
propagate along the principal axis $z$ of a chiral nanotube with $\nu = 1$
and consider the interband optical transitions
$(v,0,K) \rightarrow$ $(c,1,K)$, $(v,-1,K') \rightarrow$ $(c,0,K')$
allowed for the $\sigma_+$ polarization and $(v,1,K) \rightarrow$ $(c,0,K)$,
$(v,0,K) \rightarrow$ $(c,-1,K)$ allowed for the $\sigma_-$ polarization.
The energy conservation law reads
$E_c(n_f, k_z + q; K) -$  $E_v(n_i, k_z; K) = \hbar \omega$,
$E_c(n_f, k'_z + q; K') -$  $E_v(n_i, k'_z; K') = \hbar \omega$.
Due to the presence of the light wavevector $q$ in these equations
and the linear-in-$k_z$ terms in the energy dispersion given by
Eq.~(\ref{021}), the absorption probability rates $W(\sigma_{\pm})$
for the $\sigma_{\pm}$ polarization differ and the relative difference is
given by
\begin{equation} \label{dichr}
\frac{W(\sigma_+) - W(\sigma_-)}{W(\sigma_+) + W(\sigma_-)} \sim 
\frac{\beta q}{\hbar \omega - \Delta_{10}^{cv} }  \:,
\end{equation}
where $\Delta_{10}^{cv}$ is the band gap between $(c,1,K)$ and $(v,0,K)$
subbands.
For comparison we present also an estimation for the circular dichroism
due to the Faraday effect in the magnetic field ${\bf B} \parallel z$:
\begin{equation} \label{farad}
\frac{W(\sigma_+) - W(\sigma_-)}{W(\sigma_+) + W(\sigma_-)} \sim 
\frac{\Delta_{10}^{cv} }{\hbar \omega - \Delta_{10}^{cv} } \frac{\Phi}{\Phi_0} \:.
\end{equation}
If the condition (\ref{depol}) is not satisfied the depolarization
effect can substantially renormalize the dichroism, however the ratio
of the relative differences given by Eqs.~(\ref{dichr},\ref{farad}) holds
unchanged.

Let us now turn to an estimate of the magnitude of the magneto-chiral
dichroism which manifests itself in 
the dependence of the light absorption coefficient on  
 the product $q_{\alpha} B_{\beta}$. Here $q_{\alpha}$ and
$B_{\beta}$ are components of the light wavevector {\bf q} and the
magnetic field {\bf B} \cite{zakharchenya,enh}, (see also \cite{rikken}
and
references therein). In chiral CNs  the magneto-chiral dichroism can
be described by the following contribution to the absorption or
emission probability rate \cite{rikken}
\begin{equation}
\delta W =  B_z q_z \:[Q_{\parallel} |e_z|^2 +
Q_{\perp} (|e_x|^2 + |e_y|^2)] \:.
\end{equation}
For the interband transitions $(v, 0 , s) \rightarrow (c , 0, s)$ ($s=K,K'$)
one has $Q_{\parallel} \neq 0$, $Q_{\perp} = 0$ and $\delta W_{00}$ $\propto
B_z q_z |e_z|^2$ $= B_z \cos{\theta_0} \sin^2{\theta_0}$, where
 $\theta_0$ is the angle between the nanotube axis $z$ and the 
propagation direction of the light linearly polarized in the plane containing
the vector ${\bf q}_0$ and the axis $z$. For the transitions
$(v, 0) \rightarrow (c , 0)$ in one valley, say the valley $K$,
the ratio $\delta W_{00} / W_{00}$ is given by $\beta q_z/(\hbar \omega
- \Delta_{00})$ similarly to the previous consideration. At zero magnetic
field the transitions in the valley $K'$ lead to a contribution to
$\delta W_{00}$ of the opposite sign and the net value of
$\delta W_{00}$ vanishes.
An external magnetic field breaks the compensation, and the imbalance
in contributions from the $K$ and $K'$ valleys is governed by
a value of $[\Delta_{00} / (\hbar \omega - \Delta_{00})]$$(\Phi / \Phi_0)$.
As a result we obtain
\begin{equation}
\frac{\delta W_{00}}{W_{00}} \sim \frac{\beta q_z \Delta_{00}}{(\hbar
\omega - \Delta_{00})^2} \frac{\Phi}{\Phi_0} \:.
\end{equation}

\section{conclusion}
\label{sec:5}
All chirality induced effects considered above are not specific for chiral
CN. In principle they should be present in any mesoscopic system lacking
the mirror-reflection symmetry. In particular they should be present in
mesoscopic disordered metallic samples where all possible symmetries are
broken. Some of these effects have been already discussed 
\cite{KhmelnitskiiFalko,SpivakZhouBealMonod}.

Another system where the effects considered above can manifest
themselves
is 
a DNA molecule, which electron transport properties have been investigated 
recently \cite{helene,Schoenenberger}. Although usually DNA have large
resistance, we would like to mention in this connection that the
photoinduced electron transport effects can be measured even in diamond
\cite{sturman}. 

The photogalvanic effects should also exist in Josephson junctions with
no mirror-reflection symmetry. For example, it could be
either superconductor-carbon nanotube-superconductor 
or superconductor-disordered metal-superconductor junctions. 
The qualitative difference of the photogalvanic effects in metals is that
the circularly polarized light will induce both the normal and superfluid
components of the current through the junctions.

The theory presented above neglects effects of electron-electron
interaction including the excitonic and Luttinger-liquid
effects. In this respect we would like to mention that
$\Phi$-dependent terms
in the electron spectrum will suppress the dip in
the energy dependence of the density of states for
interacting electrons in one dimension \cite{cobden}. In disordered
samples the
 magnitude of this
effect should
match results 
\cite{altshuleraronov,nazarov,levitov} for diffusive quasi-one-dimensional
conductors.

Finally we would like to mention an effect which is inverse
to the circular and linear photogalvanic effects. Namely, in the absence
of magnetic field the {\it dc} electric current in a CN should induce
the circular polarization of the photoluminescence. In the presence of an
external magnetic field parallel to a CN, the electric current $j_z$
should
induce a change in the photoluminescence intensity proportional to
$j_{z}B_{z}$.

The work of B.Spivak was supported by Division of Material Sciences,
U.S. National  science foundation under contract No. DMR-9205144. 
The work of E.L.Ivchenko was supported by
 programmes of Russian Academy and Ministry of Sciences.
We would like to acknowledge useful discussions with D. Cobden, C. Kane,
L. Levitov and C. Markus.

\newpage

\begin{figure}
  \centerline{\epsfxsize=10cm \epsfbox{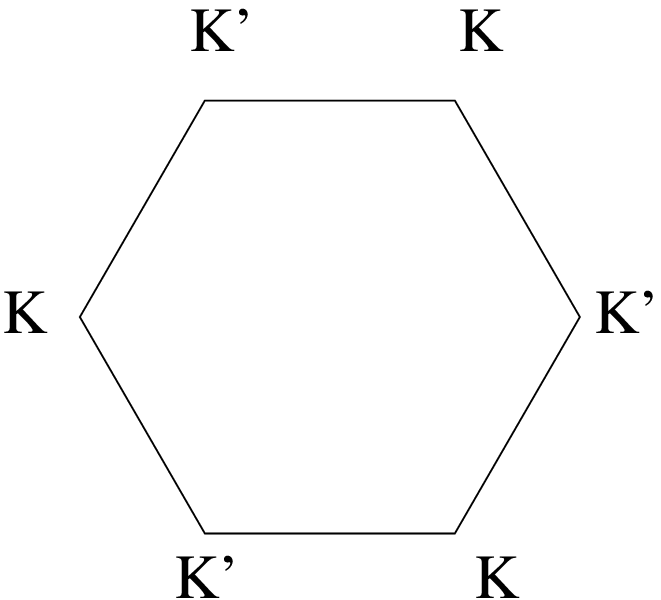}}
  \caption{
Two dimensional Brillouin zone of graphene.}
  \label{fig:fig1}
\end{figure}

\newpage

\begin{figure}
  \centerline{\epsfxsize=15cm \epsfbox{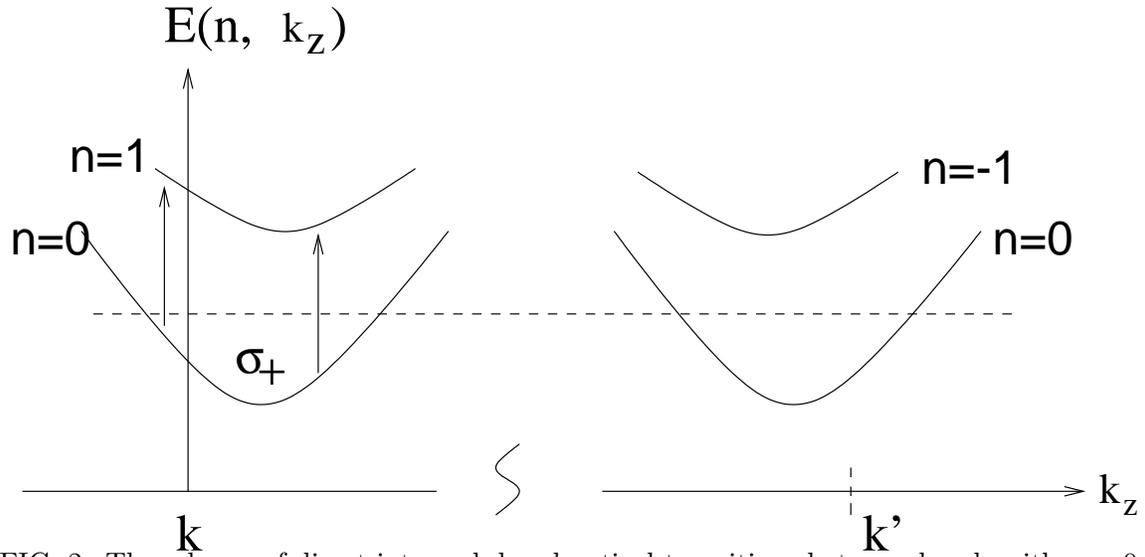}}
  \caption{The scheme of direct inter-sub-band optical transitions
between bands with $n=0$ and $n=1$. The dashed line corresponds to the 
level 
of the chemical potential in the conduction band.} \
  \label{fig:fig2}
\end{figure}

\newpage

\begin{figure}
  \centerline{\epsfxsize=10cm \epsfbox{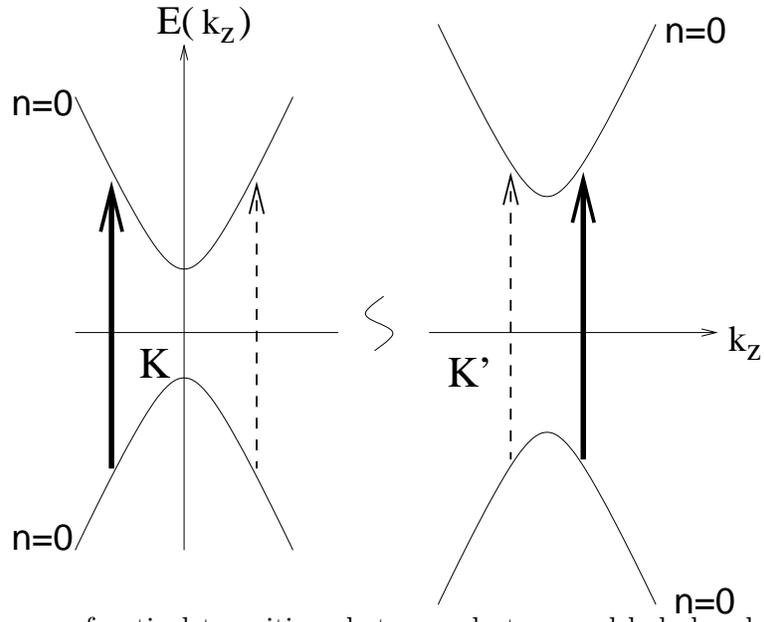}}
  \caption{ A scheme of optical transitions 
 between electron and hole bands with $n=0$ in the case of linearly
polarized light and in the presence of the external magnetic field.
The dashed lines with different width correspond to transitions with
different probabilities.}
  \label{fig:fig3}
\end{figure}

\end{document}